\documentclass[twocolumn,superscriptaddress,aps,pre]{revtex4-2}

\makeatletter
\newcommand*{\rom}[1]{\expandafter\@slowromancap\romannumeral #1@}
\makeatother

\usepackage{color}
\usepackage{graphicx}
\usepackage{hyperref}
\usepackage{amsmath,amssymb}
\usepackage{epstopdf}
\usepackage{subcaption}
\usepackage{float}

\graphicspath{{figs/}}

\usepackage{xcolor}

\begin{document}
\pagestyle{plain}
	
\title{Pseudo-Distributions: Thermodynamic Geometry and an Empirical Application}
\author{Negin Bayrami}
\affiliation{Department of Physics, University of Mohaghegh Ardabili, P.O. Box 179, Ardabil, Iran}
\author{Hosein Mohammadzadeh}
\email{mohammadzadeh@uma.ac.ir}
\affiliation{Department of Physics, University of Mohaghegh Ardabili, P.O. Box 179, Ardabil, Iran}
\author{Hamzeh Agahi}
\affiliation{Department of Mathematics, Faculty of Basic Science, Babol Noshirvani University of Technology, Shariati Ave., Babol 47148-71167, Iran }
\author{Hossein Mehri-Dehnavi}
\affiliation{Department of Physics, Faculty of Basic Science, Babol Noshirvani University of Technology, Shariati Ave., Babol 47148-71167, Iran }
\affiliation{Eigen FinTech Inc., 18 Bingham Street, Richmond Hill, ON L4C 9R1, Canada}
\author{Zahra Ebadi}
\affiliation{Department of Physics, University of Mohaghegh Ardabili, P.O. Box 179, Ardabil, Iran}

\begin{abstract}
We develop a consistent pseudo-analytic framework based on $g$-calculus for constructing deformed statistical distributions. By mapping standard algebraic operations through a monotone generator function, we systematically derive the associated pseudo-logarithmic and pseudo-exponential structures. Applying this formalism, we introduce a new family of pseudo-distributions that generalizes nonextensive statistical mechanics at both the probability density and cumulative distribution levels, recovering classical and standard nonextensive statistics as limiting cases. We investigate the thermodynamic geometry of the proposed models using the Ruppeiner metric on the equilibrium manifold. A perturbative analysis around the classical limit reveals that, to leading order, the thermodynamic scalar curvature is governed solely by the generator deformation parameter, while the nonextensivity parameter remains decoupled. To evaluate the empirical robustness of the framework, we apply the model to analyze the absolute deviations of daily West Texas Intermediate crude oil prices from their hundred-day moving average. Model comparison based on information criteria demonstrates that the proposed pseudo-distributions provide a superior description of these high-frequency financial fluctuations and their heavy-tailed characteristics compared to standard benchmarks. These results suggest that $g$-calculus offers a flexible and physically grounded mathematical tool for generating deformed statistics and analyzing their geometric properties.
\end{abstract}

\maketitle

\section{Introduction}
Generalized forms of statistical mechanics have attracted sustained attention over the past decades, largely because many complex systems cannot be satisfactorily described within the standard Boltzmann--Gibbs framework~\cite{tsallis1988possible,abe2001nonextensive}. In a wide variety of physical, biological, informational, and socioeconomic systems, one encounters long-range correlations, memory effects, anomalous transport, multifractality, hierarchical organization, or effective interactions that make the conventional exponential distributions insufficient~\cite{metzler2000random, bunde2012fractals}. This has motivated the development of a broad family of generalized statistics, each arising from a different mathematical or physical mechanism of deformation~\cite{gell2004nonextensive}.

One important class of generalizations originates from modified counting rules and generalized exclusion principles~\cite{haldane1991fractional}. A prominent example is Haldane fractional exclusion statistics, in which the dimensionality of the one-particle Hilbert space effectively depends on particle occupation, interpolating between bosonic and fermionic behavior through a statistical parameter~\cite{haldane1991fractional, wu1994statistical}. Related developments include Polychronakos-type fractional statistics, where generalized occupancy rules lead to alternative interpolation schemes between the standard quantum distributions~\cite{polychronakos1996probabilities, isakov1994statistical}. These approaches provide effective descriptions of systems whose statistical behavior cannot be reduced to the ordinary Bose--Einstein or Fermi--Dirac paradigms~\cite{wu1994statistical, de1995one}.

A different route to generalized statistics emerges from exchange properties of identical particles in reduced dimensions~\cite{leinaas1977theory, wilczek1982quantum}. In two-dimensional systems, particles known as anyons obey braid-group statistics rather than the usual permutation-group statistics, and their exchange can produce phases that continuously interpolate between bosons and fermions~\cite{wilczek1982quantum, khare2005fractional}. The resulting fractional exchange statistics has played a central role in the study of low-dimensional quantum systems, topological phases of matter, and quasiparticle excitations such as those appearing in the fractional quantum Hall effect~\cite{halperin1984statistics, wilczek1990fractional}. In such cases, the deformation of statistical behavior is rooted in topology and exchange structure rather than in energetic or entropic modifications alone~\cite{leinaas1977theory, khare2005fractional}.

Generalized statistics can also arise from deformed algebraic frameworks~\cite{vg1990quantum}. In this direction, various \(q\)-deformed oscillator algebras, quantum groups, and nonlinear commutation relations have been used to construct modified distribution functions and thermodynamic relations~\cite{biedenharn1989quantum, macfarlane1989q, sun1989q, arik1976hilbert}. Here the deviation from standard statistics is encoded directly in the algebraic structure of creation and annihilation operators or, more generally, in deformed composition laws~\cite{macfarlane1989q, chakrabarti1991p}. These algebraic approaches are especially useful for describing systems in which nontrivial symmetry structures or effective microscopic constraints alter the conventional thermodynamic description~\cite{vg1990quantum}.

Among the many families of generalized statistics, a particularly influential class is formed by nonextensive statistics~\cite{tsallis1988possible, tsallis2009introduction}. In this setting, the standard additive entropy is replaced by generalized entropic functionals, leading to deformed logarithmic and exponential functions in the corresponding equilibrium distributions~\cite{tsallis1988possible, abe2001nonextensive}. The Tsallis formalism is the best-known representative of this class and has been widely studied in the context of complex systems with long-range interactions, scale-invariant behavior, and non-Markovian effects~\cite{tsallis2009introduction}. Closely related to this line of research is the Kaniadakis formalism, which introduces another generalized logarithm and exponential with distinct symmetry and relativistic-motivated features~\cite{kaniadakis2001non, kaniadakis2002statistical}. Although these frameworks differ in construction and interpretation, both share a fundamental structural aspect: the ordinary exponential function appearing in classical statistical distributions is replaced by a generalized exponential~\cite{kaniadakis2013theoretical, scarfone2006canonical}.

This observation suggests a broader perspective: many generalized statistical frameworks may be examined through deformations of the mathematical structure underlying the exponential distribution itself~\cite{tsallis1988possible, tsallis2009introduction, kaniadakis2001non}. From this perspective, the central object is no longer only the entropy functional, but also the class of generalized exponential-type functions from which the associated probability distributions are built. Such a viewpoint has recently motivated further extensions beyond the standard nonextensive setting.

In particular, generalized distribution functions based on the Mittag--Leffler function have recently been proposed. Within this construction, one obtains classical Mittag--Leffler Maxwell--Boltzmann distributions as well as quantum Mittag--Leffler Bose--Einstein and Fermi--Dirac distributions. Because the Mittag--Leffler function naturally appears in fractional calculus, anomalous relaxation, and memory-dependent processes, these distributions provide a promising framework for systems whose statistical behavior reflects nonlocality in time, fractional dynamics, or complex relaxation patterns. Their introduction further illustrates that generalized statistics may emerge not only from entropy deformation or exclusion rules, but also from replacing the ordinary exponential kernel with a richer special-function structure~\cite{seifi2025intrinsic, seifi2025mittag}.

Another recent direction is based on nonlinear measures and nonadditive integration schemes. In particular, it has been shown that the use of nonlinear capacities associated with the Choquet integral allows one to construct generalized exponential-type distributions, including both classical and quantum forms. In this approach, the deformation is encoded in the underlying measure structure rather than solely in the entropy or algebraic relations. This provides a conceptually distinct mechanism for generating effective statistical models, especially for systems in which interactions, uncertainty, or collective effects are more naturally represented through nonadditive set functions~\cite{choquet1954theory, bayrami2026quantum}.

These examples show that generalized statistics do not arise from a unique source. Rather, they form a broad landscape of theories whose differences reflect distinct physical intuitions and mathematical constructions: generalized exclusion~\cite{haldane1991fractional, wu1994statistical}, exchange topology~\cite{leinaas1977theory, wilczek1982quantum}, deformed algebra~\cite{vg1990quantum}, nonextensive entropy~\cite{tsallis1988possible, kaniadakis2001non}, fractional-function kernels~\cite{podlubny1998fractional}, and nonlinear integration~\cite{choquet1954theory, denneberg1994non}. A natural question therefore arises: are there alternative mathematical frameworks capable of systematically generating generalized statistical distributions, while also allowing one to investigate their structural and thermodynamic properties?

The aim of the present work is to explore such an alternative route. Instead of deforming the entropy functional, the occupation rule, or the integration measure, we adopt a pseudo-analytic framework based on \(g\)-calculus~\cite{pap2008generalized}. In this approach, the deformation is introduced at the level of the underlying algebraic and differential operations, leading naturally to generalized logarithmic, exponential, and distribution functions. This viewpoint provides a flexible formalism for constructing families of pseudo-deformed distributions and for examining their relation to known generalized statistics.

Within this framework, we first develop the pseudo-analytic formalism based on g-calculus and introduce pseudo-generalizations of the exponential distribution constructed at both the PDF and CDF levels. Building on these results, we formulate a generalized pseudo-Tsallis distribution and investigate its analytical properties. We then examine its thermodynamic geometry by performing a perturbative expansion around the classical Maxwell–Boltzmann limit. The analysis reveals that, to first order around $q=1$, the thermodynamic curvature is governed entirely by the deformation parameter $\alpha$, whereas the nonextensivity parameter $q$ does not contribute to the curvature. Finally, the proposed pseudo-Tsallis distribution is applied to a real WTI crude oil dataset, where its performance is assessed through standard model selection criteria.

The rest of the paper is organized as follows. Section 2 introduces the basic concepts of g-calculus and the pseudo-analytic framework. Section 3 presents pseudo-generalizations of the exponential distribution at the PDF and CDF levels, while Section 4 investigates their thermodynamic geometry. Section 5 develops the pseudo-Tsallis distribution, and Section 6 studies its associated thermodynamic geometry. Section 7 briefly reviews the statistical performance of the pseudo-exponential model, whereas Section 8 presents the application of the pseudo-Tsallis distribution to the WTI crude oil dataset. Finally, Section 9 summarizes the main findings and discusses their physical implications.

\section{Pseudo-operations based on $g$-calculus}

A fundamental class of pseudo-operations can be systematically constructed within the framework of $g$-calculus, as introduced by Pap \cite{pap2008generalized}. This formalism offers a robust method for generating generalized algebraic and functional structures by transporting standard operations through a monotone and continuous function. We consider a generator function $g$ such that
\[
g:[a,b]\to [-\infty,\infty]
\qquad \text{or} \qquad
g:[a,b]\to [0,\infty],
\]
where $[a,b]$ represents a closed or semi-closed subinterval of the extended real line $\bar{\mathbb{R}}=[-\infty,\infty]$. By assuming that $g$ is monotone and continuous, its inverse $g^{-1}$ is well-defined over the range of $g$.

The core principle of $g$-calculus lies in defining generalized functions and operations on the original space by conjugating ordinary functions with the mapping $g$. This approach allows many classical mathematical objects, such as logarithmic and exponential functions, to be reformulated in a deformed setting. Specifically, for any function $f$ defined on the range of $g$, we can associate a corresponding \emph{$g$-function}, denoted by $f_g(x)$, through the relation:
\begin{equation}
f_g(x)=g^{-1}\!\left(f(g(x))\right).
\label{122}
\end{equation}
This construction can be interpreted as a similarity transformation of the function $f$ via the generator $g$. Such a transformation preserves the underlying structural role of $f$ while adapting its behavior to a generalized domain dictated by the choice of $g$.

Particular interest lies in cases where $f$ corresponds to the standard logarithm or exponential functions. By applying the transformation in \eqref{122}, we obtain the \emph{$g$-logarithm}:
\begin{equation}
\ln_g(x)=g^{-1}\!\left(\ln(g(x))\right),
\end{equation}
and the \emph{$g$-exponential}:
\begin{equation}
\exp_g(x)=g^{-1}\!\left(\exp(g(x))\right).
\end{equation}
It is straightforward to see that these generalized functions recover the standard logarithm and exponential in the limit where $g(x)=x$, thereby ensuring that the $g$-formalism contains the ordinary statistics as a particular case.

Building upon this idea, the $g$-calculus framework can be integrated with the $q$-deformed functions prevalent in nonextensive statistics. We recall that the standard $q$-logarithm and $q$-exponential are defined, respectively, by \cite{tsallis1988possible}:
\begin{equation}
\ln_q(x)=\frac{x^{1-q}-1}{1-q},
\qquad x>0,\quad q\neq 1,
\end{equation}
and
\begin{equation}
\exp_q(x)=\left[1+(1-q)x\right]_+^{\frac{1}{1-q}},
\qquad q\neq 1,
\end{equation}
where $[A]_+=\max\{A,0\}$. In the limit $q\to 1$, these expressions reduce to the classical logarithm and exponential functions. By combining these two layers of deformation, we can define the \emph{$g$-$q$-logarithm} and the \emph{$g$-$q$-exponential} as:
\begin{align}
\ln_{g,q}(x) &= g^{-1}\!\left(\ln_q(g(x))\right), \\
\exp_{g,q}(x) &= g^{-1}\!\left(\exp_q(g(x))\right).
\end{align}
These functions represent a broader class of deformed mappings where the departure from standard behavior arises simultaneously from the nonextensivity parameter $q$ and the structural generator $g$. Consequently, they provide a flexible framework for constructing generalized statistical distributions that extend beyond the standard Tsallis form.

It is important to note that, provided the relevant domains are well-defined, the functions $\ln_{g,q}$ and $\exp_{g,q}$ maintain their reciprocal relationship:
\begin{align}
\exp_{g,q}\!\left(\ln_{g,q}(x)\right) &= x,
\qquad \text{for } g(x)>0,
\\
\ln_{g,q}\!\left(\exp_{g,q}(x)\right) &= x,
\qquad \text{for } 0<\exp_q(g(x))<\infty.
\end{align}
Thus, the pair $(\ln_{g,q},\exp_{g,q})$ fulfills the same structural and functional role in the generalized setting as the ordinary logarithm and exponential do in standard analysis, or as the pair $(\ln_q,\exp_q)$ does within the Tsallis framework.

\section{Generalizations of the Exponential distribution by $g$-Calculus}

The exponential distribution is one of the simplest and most widely used probability distributions in statistics. It is widely applied in lifetime data analysis and also appears naturally in statistical physics through the Boltzmann--Gibbs formalism. A non-negative random variable $X$ is said to have an exponential distribution if its probability density function is given by
\begin{equation}
f_X(x)=\lambda e^{-\lambda x}, \qquad x>0,\ \lambda>0.
\end{equation}

The exponential distribution has been used in various areas such as econophysics~\cite{mantegna2000introduction},  condensed matter physics \cite{bernasconi1979anomalous,kakalios1987stretched,macdonald1985frequency}, theoretical physics \cite{budiyono2013quantization,druagulescu2001exponential}, astronomy and astrophysics \cite{collier2004magnetospheric}, neuroscience \cite{zeman2015exponential,trappenberg2009fundamentals}, mechanical systems \cite{granato1956theory}, and climate science \cite{field2005parametrization}. Because of its mathematical simplicity and broad applicability, the exponential distribution has been generalized in different directions.
In this section, some generalizations of the exponential distribution based on the framework of $g$-calculus are discussed.

\subsection{Pseudo-generalization at the PDF level}
We first consider the pseudo-generalization of the exponential distribution at the level of the probability density function (PDF)~\cite{mehri2019pseudo}. Starting from the exponential density and using the class of $g$-functions defined by
\[
g(x)=x^\alpha,\qquad \alpha>0,
\]
together with the transformation
\[
g^{-1}\!\big(f_X(g(x))\big),
\]
we obtain the following pseudo-exponential probability density function:
\begin{equation}
f^{(\mathrm{PDF})}_{X,\alpha,\lambda}(x)=
\lambda^{1/\alpha}\frac{e^{-\lambda x^\alpha}}
{\Gamma\left(1+\frac{1}{\alpha}\right)},
\quad
x>0,\quad \alpha>0,\quad \lambda>0.
\end{equation}
This distribution reduces to the standard exponential density when \(\alpha=1\).

\subsection{Pseudo-generalization at the CDF level}

We next consider an alternative pseudo-generalization constructed at the level of the cumulative distribution function (CDF). Let
\[
F_X(x)=1-e^{-\lambda x},
\qquad x>0,\quad \lambda>0,
\]
be the cumulative distribution function of the exponential distribution. By applying the same \(g\)-deformation with \(g(x)=x^\alpha\), we are led to the pseudo-generalized CDF
\begin{equation}
F_{X,\alpha,\lambda}(x)=
\left(1-e^{-\lambda x^\alpha}\right)^{1/\alpha},\quad
x>0,\quad \lambda>0,\quad \alpha>0.
\end{equation}

Differentiating this expression with respect to \(x\), the corresponding probability density function is obtained as
\begin{equation}
\begin{aligned}
f^{(\mathrm{CDF})}_{X,\alpha,\lambda}(x)
&=
\lambda x^{\alpha-1}e^{-\lambda x^\alpha}
\left(1-e^{-\lambda x^\alpha}\right)^{\frac{1-\alpha}{\alpha}},\\
&\quad x>0,\quad \lambda>0,\quad \alpha>0.
\end{aligned}
\end{equation}

Again, for \(\alpha=1\), this construction recovers the standard exponential distribution.

\section{Thermodynamic geometry of the Pseudo-Exponential distribution
}

In the preceding sections of this study, we established a rigorous mathematical framework for the extension of Exponential and Tsallis statistics through the application of $g$-calculus and pseudo-analysis. While the mathematical consistency of these generalized distributions has been demonstrated, a fundamental question remains: what are the concrete physical implications of such deformations on  thermodynamic systems? To address this, we shift our focus from pure mathematical formalism to a physical diagnostic approach using the framework of thermodynamic geometry.

To investigate the geometric properties of the system, we employ the framework of thermodynamic geometry. Consider an ideal gas in $D$ spatial dimensions confined in a volume $L^D$ with a general dispersion relation $\epsilon = a p^\sigma$. In the thermodynamic limit, the total particle number and internal energy can be written in terms of the occupation number $n(\epsilon)$ and the single–particle density of states $\Omega(\epsilon)$ as
\begin{eqnarray}
N &=& \int_{0}^{\infty} n(\epsilon)\,\Omega(\epsilon)\,d\epsilon, \\
U &=& \int_{0}^{\infty} \epsilon\, n(\epsilon)\,\Omega(\epsilon)\,d\epsilon .
\end{eqnarray}
The single-particle density of states is given by $\Omega(\epsilon)=\frac{A^D}{\Gamma(D/2)}\,\epsilon^{\frac{D}{\sigma}-1}$, where $A$ encapsulates the volume and dispersion constants.
The framework of thermodynamic geometry, as pioneered by Ruppeiner and Weinhold~\cite{ruppeiner1979thermodynamics, weinhold1975additional}, treats the space of thermodynamic equilibrium states as a Riemannian manifold. To evaluate the geometry, the metric of the thermodynamic state space can be obtained from the Hessian of the logarithm of the partition function. In the canonical representation the metric components are defined as~\cite{janyszek1990riemannian}
\begin{equation}
g_{ij}=\frac{\partial^2 \ln Z}{\partial \beta^i \partial \beta^j},
\end{equation}
where $(\beta^1,\beta^2)=(\beta,\gamma)$, with $\beta=1/(kT)$ and $\gamma=-\mu/(kT)$~\cite{mirza2010thermodynamic}. The components of the metric are given by

\begin{eqnarray}
\begin{aligned}
g_{\beta\beta} &=& -\left(\frac{\partial U}{\partial \beta}\right)_\gamma, \\
g_{\beta\gamma} &=& -\left(\frac{\partial U}{\partial \gamma}\right)_\beta, \\
g_{\gamma\gamma} &=& -\left(\frac{\partial N}{\partial \gamma}\right)_\beta .
\end{aligned}
\end{eqnarray}

The central quantity of interest in this theory is the thermodynamic scalar curvature, $R$

\begin{equation}
R = -\frac{
	\begin{vmatrix}
	g_{\beta\beta} & g_{\beta\gamma} & g_{\gamma\gamma} \\
	g_{\beta\beta,\beta} & g_{\beta\gamma,\beta} & g_{\gamma\gamma,\beta} \\
	g_{\beta\beta,\gamma} & g_{\beta\gamma,\gamma} & g_{\gamma\gamma,\gamma}
	\end{vmatrix}
}{
	2
	\begin{vmatrix}
	g_{\beta\beta} & g_{\beta\gamma} \\
	g_{\gamma\beta} & g_{\gamma\gamma}
	\end{vmatrix}^{2}
}.
\end{equation}

This curvature is not merely a geometric abstraction but a direct measure of the inherent statistical correlations and microscopic interactions within the system. From the established literature on quantum gases, we know that:
\begin{itemize}
	\item For an ideal classical gas obeying Maxwell--Boltzmann statistics, where particles are non-interacting and independent, the thermodynamic curvature vanishes identically ($R = 0$).
	\item For a quantum Bose gas, the inherent tendency of particles to occupy the same state (statistical attraction) manifests as a positive scalar curvature ($R > 0$).
	\item For a quantum Fermi gas, where the Pauli exclusion principle prevents particles from occupying the same state (statistical repulsion), the curvature is negative ($R < 0$).
\end{itemize}
Thus, by calculating $R$ for our $g$-deformed distributions, we can identify whether the pseudo-analysis induces an effective "attractive" or "repulsive" statistical behavior, effectively mapping the mathematical parameter $\alpha$ to a physical interaction regime.


We first consider the thermodynamic geometry associated with the PDF-level pseudo-exponential construction introduced in the preceding section. The $g$-deformed PDF can be expanded around the classical limit, where the deformation parameter $\alpha$ approaches unity. By keeping terms up to the first order in $(\alpha-1)$, we obtain the following approximation:
\begin{equation}
f(x) \simeq e^{-x} \left[ 1 + (\alpha-1) \left( 1 - \gamma_{E} - x \ln x \right) \right].
\label{equ:expanded_pdf}
\end{equation}
Here, $x = \beta(\epsilon - \mu)$ represents the dimensionless energy parameter, and $\gamma_{E} \approx 0.577$ is the Euler--Mascheroni constant. Substituting this distribution into the relations for the total particle number and internal energy as the energy distribution function of the system, the corresponding thermodynamic metric elements are obtained. The expressions for the internal energy and the total particle number can be expanded in the form
\begin{equation}
X = X_{0} + (\alpha - 1)X_{1},
\end{equation}
where $X_{0}$ represents the classical Maxwell--Boltzmann limit and $X_{1}$ denotes the first--order correction associated with the pseudo-generalized parameter $\alpha$.

So, the particle number and internal energy can be written as
\begin{equation}
N = N_{0} + (\alpha-1)\,N_{1}(z,\beta),
\end{equation}

\begin{equation}
U = U_{0} + (\alpha-1)\,U_{1}(z,\beta),
\end{equation}
where
\begin{equation*}
N_{0}=\frac{\sqrt{\pi}\,z}{2\beta^{3/2}}, 
\qquad
U_{0}=\frac{3\sqrt{\pi}\,z}{4\,\beta^{5/2}},
\end{equation*}

Using the metric tensor, the thermodynamic scalar curvature $R$ is calculated up to first order in $(\alpha-1)$.
\begin{equation}
R = R_{0} + (\alpha-1)\,R_{1}(z,\beta).
\end{equation}
Where $R_{0}=0$.
The functions $N_{1}(z,\beta)$, $U_{1}(z,\beta)$ and $R_{1}(z,\beta)$ are given explicitly in Appendix.

In contrast to the previous work, where the pseudo-generalization was introduced at the level of the probability density function, here we consider an alternative construction obtained by generalizing the cumulative distribution function. This procedure leads to a different pseudo-generalized exponential distribution, which we analyze within the same thermodynamic geometric framework presented earlier.

To analyze the deviation from the classical exponential case, we expand the PDF around $\alpha=1$. Retaining terms up to first order in $(\alpha-1)$ gives
\begin{equation}
f(x)\simeq
e^{-x}
\left[
1+(\alpha-1)
\left(
\ln x-x\ln x-\ln\left(1-e^{-x}\right)
\right)
\right].
\label{equ:cdf}
\end{equation}

\begin{figure*}[htbp]
	\centering
	\includegraphics[width=.75\textwidth]{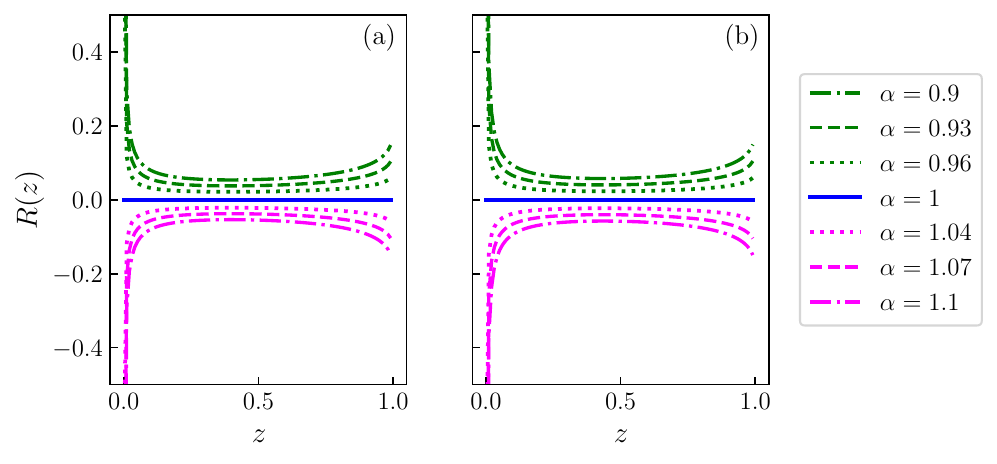}
	\caption{Thermodynamic scalar curvature $R(z)$ as a function of the fugacity $z$ for different values of the pseudo-generalization parameter $\alpha$. Panel (a) corresponds to the pseudo-generalized exponential model obtained directly from the PDF, while panel (b) represents the model constructed by pseudo-generalizing the CDF. In both cases, $\alpha<1$ leads to positive curvature, $\alpha>1$ results in negative curvature, and the classical exponential limit $\alpha=1$ yields vanishing curvature.}
	\label{fig:RPScombined}
\end{figure*}

This approximate distribution is then employed within the thermodynamic framework introduced earlier to analyze the associated thermodynamic curvature. Similarly to the previous subsection, for the total particle number and internal energy, we have
\begin{equation*}
N_{0}=\frac{\sqrt{\pi}\,z}{2\beta^{3/2}},
\qquad
U_{0}=\frac{3\sqrt{\pi}\,z}{4\,\beta^{5/2}}.
\end{equation*}
As before, the functions $N_{1}(z,\beta)$, $U_{1}(z,\beta)$, and $R_{1}(z,\beta)$ are given explicitly in the Appendix A.
The behavior of the resulting curvature as a function of the fugacity parameter $z$ is illustrated in Fig.~\ref{fig:RPScombined}.

Figure~\ref{fig:RPScombined} exhibits a high degree of similarity, rendering the choice between the two expanded relations inconsequential. In both scenarios, the parameter $\alpha$ acts as a critical threshold: the system separates into positive and negative regimes depending on whether $\alpha$ is less than or greater than unity. Consequently, the underlying dynamics reduce to either attractive or repulsive interactions within the system.

Moreover, it makes no difference whether one first considers the pseudo-exponential cumulative distribution function, derives the probability density function from it, and then expands the resulting expression, or instead directly pseudo-deforms the PDF and subsequently applies the expansion. In both approaches, the resulting expanded functions exhibit nearly identical behavior and therefore induce the same physical characteristics in the system. To first order in the deformation, the regime $\alpha<1$ is associated predominantly with attractive behavior among the particles, whereas the regime $\alpha>1$ corresponds to predominantly repulsive behavior. In the limit $\alpha=1$, the Maxwell--Boltzmann distribution is recovered, corresponding to a non-interacting system. In general, the parameter $\alpha$ serves as a global control parameter in the thermodynamic geometry of the system.

\section{Generalizations of the Nonextensive Tsallis distribution by $g$-Calculus}
\label{101}

Tsallis statistics \cite{tsallis1988possible,gell2004nonextensive} is one of the best-known frameworks in nonextensive statistical mechanics for the study of complex systems. The corresponding $q$-exponential and $q$-logarithm functions are defined by \cite{tsallis1988possible,gell2004nonextensive}
\begin{equation}
\exp_q(x)=
\begin{cases}
\exp(x), & q=1,\\[4pt]
\left[1+(1-q)x\right]^{\frac{1}{1-q}}, & q\neq 1,\; 1+(1-q)x\ge 0,\\[4pt]
0, & q\neq 1,\; 1+(1-q)x<0,
\end{cases}
\end{equation}
and
\begin{equation}
\ln_q(x)=
\begin{cases}
\ln(x), & q=1,\; x>0,\\[4pt]
\dfrac{x^{1-q}-1}{1-q}, & q\neq 1,\; x>0,\\[6pt]
\text{undefined}, & x\le 0.
\end{cases}
\end{equation}
The following properties for the $q$-exponential can be proven%
\begin{eqnarray}
\begin{aligned}
\exp _{q}\left( \ln _{q\text{ }}\left( x\right) \right) &=1\text{ for }x>0,
\\
\ln _{q}\left( \exp _{q\text{ }}\left( x\right) \right) &=1\text{ for }%
0<\exp _{q\text{ }}\left( x\right) <+\infty ,
\end{aligned}
\end{eqnarray}

Tsallis-type distributions play an important role in the description of systems with long-range interactions, memory effects, and nonlocal correlations. They have been applied in many areas of science, including economics \cite{borland2002option,ludescher2011universal}, earthquakes \cite{antonopoulos2014evidence}, quantum physics \cite{caruso2008nonadditive,nobre2011nonlinear}, plasma physics \cite{liu2008superdiffusion}, quantum thermodynamics \cite{abe2003validity}, trapped atoms \cite{devoe2009power}, cold atom systems \cite{douglas2006tunable,lutz2013beyond}, and thermodynamic geometry \cite{adli2019condensation}.

In this section, we consider pseudo-generalizations of the $q$-Tsallis exponential distribution obtained by means of the $g$-calculus framework with
\[
g(x)=x^\alpha,\qquad \alpha>0.
\]

\subsection{Pseudo-generalization at the PDF level}

Let $f_X(x)$ denote the $q$-Tsallis exponential probability density function of a non-negative continuous random variable $X$. By applying the transformation
\[
g^{-1}\!\big(f_X(g(x))\big),
\]
with \(g(x)=x^\alpha\), we obtain the following pseudo-\(q\)-Tsallis exponential probability density function.

	The pseudo-\(q\)-Tsallis exponential probability density function of \(X\) is defined by
	\begin{widetext}
	\begin{equation}
	f^{(\mathrm{PDF})}_{X,\lambda,q,\alpha}(x)=
	\begin{cases}
	\left((2-q)\lambda\left(1+(q-1)\lambda x^\alpha\right)^{\frac{1}{1-q}}\right)^{1/\alpha},
	& 1<q<2,\; x>0,\; \lambda>0,\; \alpha>0,\\[8pt]
	\left(\lambda e^{-\lambda x^\alpha}\right)^{1/\alpha},
	& q=1,\; x>0,\; \lambda>0,\; \alpha>0,\\[8pt]
	\left((2-q)\lambda\left(1+(q-1)\lambda x^\alpha\right)^{\frac{1}{1-q}}\right)^{1/\alpha},
	& q<1,\; 0<x<\dfrac{1}{\lambda(1-q)},\; \lambda>0,\; \alpha>0.
	\end{cases}
	\end{equation}
	\end{widetext}
For \(q=1\), this expression reduces to the pseudo-exponential distribution introduced in the previous section.

\subsection{Pseudo-generalization at the CDF level}

We next consider the pseudo-generalization at the level of the cumulative distribution function (CDF). Let \(F_X(x)\) denote the \(q\)-Tsallis exponential cumulative distribution function.

The pseudo-\(q\)-Tsallis exponential cumulative distribution function of \(X\) is defined by
	\begin{widetext}
	\begin{equation}
	F_{X,\lambda,q,\alpha}(x)=
	\begin{cases}
	\left[1-\left(1+(q-1)\lambda x^\alpha\right)^{\frac{q-2}{q-1}}\right]^{1/\alpha},
	& 1<q<2,\; x>0,\; \lambda>0,\; \alpha>0,\\[8pt]
	\left(1-e^{-\lambda x^\alpha}\right)^{1/\alpha},
	& q=1,\; x>0,\; \lambda>0,\; \alpha>0,\\[8pt]
	\left[1-\left(1+(q-1)\lambda x^\alpha\right)^{\frac{q-2}{q-1}}\right]^{1/\alpha},
	& q<1,\; 0<x<\dfrac{1}{\lambda(1-q)},\; \lambda>0,\; \alpha>0.
	\end{cases}
	\end{equation}
\end{widetext}

	Differentiating this expression with respect to \(x\), the corresponding probability density function is obtained as
\begin{widetext}
	\begin{equation}
	f^{(\mathrm{CDF})}_{X,\lambda,q,\alpha}(x)=
	\begin{cases}
	\lambda(2-q)x^{\alpha-1}
	\left(1+(q-1)\lambda x^\alpha\right)^{\frac{1}{1-q}}
	\left[
	1-\left(1+(q-1)\lambda x^\alpha\right)^{\frac{q-2}{q-1}}
	\right]^{\frac{1-\alpha}{\alpha}},
	&
	1<q<2,\; x>0,\; \lambda>0,\; \alpha>0,\\[8pt]
	
	\lambda x^{\alpha-1}e^{-\lambda x^\alpha}
	\left(1-e^{-\lambda x^\alpha}\right)^{\frac{1-\alpha}{\alpha}},
	&
	q=1,\; x>0,\; \lambda>0,\; \alpha>0,\\[8pt]
	
	\lambda(2-q)x^{\alpha-1}
	\left(1+(q-1)\lambda x^\alpha\right)^{\frac{1}{1-q}}
	\left[
	1-\left(1+(q-1)\lambda x^\alpha\right)^{\frac{q-2}{q-1}}
	\right]^{\frac{1-\alpha}{\alpha}},
	&
	q<1,\; 0<x<\dfrac{1}{\lambda(1-q)},\; \lambda>0,\; \alpha>0.
	\end{cases}
	\end{equation}
\end{widetext}

denoting by $X\sim PTE\left( \lambda ,q,\alpha \right) .$\newline

\begin{figure}[htbp]
\centering
\includegraphics[width=7.5 cm]{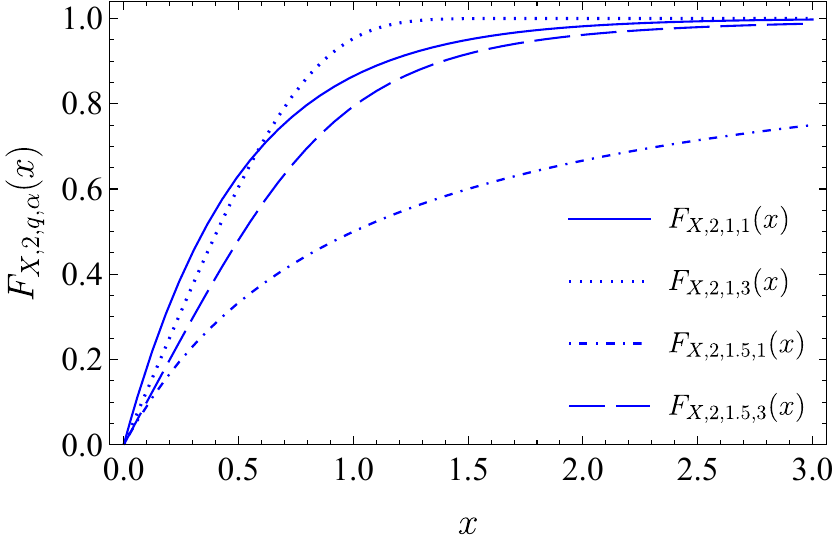}
\caption{ {Graphs of $F_{X,\protect\lambda ,q,\protect\alpha }\left(
x\right) $ for $\protect\lambda=2$ and some different values of $q$ and $%
\protect\alpha$. } }
\label{fig1}
\end{figure}

\begin{figure}[htbp]
\centering
\includegraphics[width=7.5 cm]{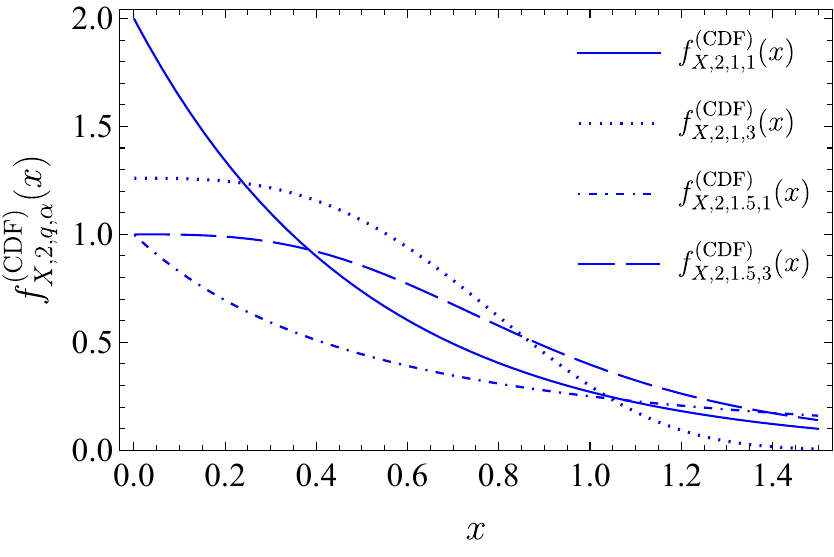}
\caption{ {Graphs of $f^{(\mathrm{CDF})}_{X,\protect\lambda ,q,\protect\alpha }\left(x\right) $ for $\protect\lambda=2$ and some different values of $q$ and $%
\protect\alpha$. } }
\label{fig2}
\end{figure}

Figures~\ref{fig1} and \ref{fig2} show the graphs of \(F_{X,\lambda,q,\alpha}(x)\) and \(f^{(\mathrm{CDF})}_{X,\lambda,q,\alpha}(x)\), respectively.

We compare the pseudo-$q$-Tsallis exponential distribution with other
distributions using AIC and BIC such that 
\begin{eqnarray}
\begin{aligned}
&&\text{AIC}&\overset{}{:=}2\kappa -2\ln \mathcal{L}, \\
&&\text{BIC}&\overset{}{:=}-2\ln \mathcal{L}+\kappa \ln n,
\end{aligned}
\end{eqnarray}%
where $n$ is the number of data, $\kappa $ is the number of estimated
parameters and $L$ is the maximum value of the likelihood function.

The maximum likelihood estimates (MLEs) of the parameters $\lambda ,q,\alpha 
$ based on an independent and identically distributed sample $x_{1},\cdots
,x_{n}\sim PTE\left( \lambda ,q,\alpha \right) $ can be obtained from

\begin{equation}
\begin{aligned}
\log &\mathcal{L}_{\mathrm{PTE}}(\lambda,q,\alpha)
=
\\[-2pt]
&\sum_{i=1}^{n}
\Bigg[
\log\lambda+\log(2-q)
+\frac{1}{1-q}
\log\!\left(1+(q-1)\lambda x_i^{\alpha}\right)
\\
&\qquad
+(\alpha-1)\log x_i
\\
&\qquad
+\frac{1-\alpha}{\alpha}
\log\!\left(
1-
\left(1+(q-1)\lambda x_i^{\alpha}\right)^{\frac{q-2}{q-1}}
\right)
\Bigg].
\end{aligned}
\end{equation}
which is solved numerically with the package \textquotedblleft
nlminb\textquotedblright\ from the statistical software R.

\section{Thermodynamic geometry of the Pseudo-Tsallis distribution}

We now consider the thermodynamic geometry associated with the pseudo-Tsallis construction introduced in the previous section. In analogy with the pseudo-exponential case, we examine the deviation from the classical limit by expanding the corresponding distribution around the undeformed point. Since the pseudo-Tsallis model depends on two deformation parameters, namely $\alpha$ and $q$, the expansion is performed around $\alpha=1$ and $q=1$, where the standard Maxwell--Boltzmann form is recovered.

We first consider the pseudo-Tsallis construction at the PDF-level. 
Expanding up to first order in $(\alpha-1)$ and $(q-1)$, we obtain
\begin{equation}
\begin{aligned}
f(x,\alpha,q)&\simeq{}
e^{-x}
\Bigg[
1+(\alpha-1)x(1-\ln x)
\\
&
+(q-1)
\left(
-1+\frac{x^{2}}{2}
\right)
\Bigg].
\label{eq:tsallis_pdf_expansion}
\end{aligned}
\end{equation}
Here, as before, $x=\beta(\epsilon-\mu)$ denotes the dimensionless energy variable. This approximation allows one to evaluate the thermodynamic quantities and the corresponding metric elements perturbatively around the classical limit.

Accordingly, the total particle number and internal energy can be written in the perturbative form
\begin{equation}
X=X_{0}+(\alpha-1)X_{\alpha}+(q-1)X_{q},
\end{equation}
where $X_{0}$ denotes the undeformed Maxwell--Boltzmann contribution, while $X_{\alpha}$ and $X_{q}$ represent the first-order corrections induced by the parameters $\alpha$ and $q$, respectively. Therefore, one may write
\begin{equation}
N=N_{0}+(\alpha-1)N_{\alpha}(z,\beta)+(q-1)N_{q}(z,\beta),
\end{equation}
\begin{equation}
U=U_{0}+(\alpha-1)U_{\alpha}(z,\beta)+(q-1)U_{q}(z,\beta).
\end{equation}
where
\begin{equation*}
N_{0}=\frac{\sqrt{\pi}\,z}{2\beta^{3/2}}, 
\qquad
U_{0}=\frac{3\sqrt{\pi}\,z}{4\,\beta^{5/2}},
\end{equation*}

Similarly, the thermodynamic scalar curvature can be expanded as
\begin{equation}
R=R_{0}+(\alpha-1)R_{\alpha}(z,\beta)+(q-1)R_{q}(z,\beta),
\end{equation}
where $R_{0}=0$ and $R_{q}=0$. The explicit expressions for $N_{\alpha}(z,\beta)$, $N_{q}(z,\beta)$, $U_{\alpha}(z,\beta)$, $U_{q}(z,\beta)$, and $R_{\alpha}(z,\beta)$the associated curvature corrections are presented in the Appendix.

We next turn to the pseudo-Tsallis construction at the CDF- level. In this case, the pseudo-generalization is introduced through the cumulative distribution function, leading to a different effective probability density.
Its first-order expansion around $\alpha=1$ and $q=1$ is

\begin{equation}
\begin{aligned}
f(x,\alpha,q)&\simeq{}
e^{-x}
\Bigg[
1+(\alpha-1)
\left(
(1-x)\ln x-\ln(1-e^{-x})
\right)
\\
&
+(q-1)
\left(
-1+\frac{x^{2}}{2}
\right)
\Bigg].
\label{eq:tsallis_cdf_expansion}
\end{aligned}
\end{equation}
Using this approximate form, the thermodynamic quantities can again be evaluated within the same geometric framework. 
As before, $N_{0}$ and $U_{0}$ take the same values.

As in the previous case, the functions $N_{\alpha}(z,\beta)$, $N_{q}(z,\beta)$, $U_{\alpha}(z,\beta)$, $U_{q}(z,\beta)$, and $R_{\alpha}(z,\beta)$ are presented explicitly in the Appendix B. 

An important consequence of the first-order expansion around $q=1$ is that the nonextensivity parameter $q$ does not contribute to the thermodynamic curvature. Consequently, to first order, the thermodynamic curvature is determined entirely by the deformation parameter $\alpha$, which governs both the magnitude and the sign of the effective statistical interaction. Specifically, for $\alpha>1$, the curvature is negative, indicating repulsive interactions, whereas for $\alpha<1$, the curvature is positive, corresponding to attractive interactions. This result suggests that the geometric structure of the system is controlled solely by the deformation parameter at leading order, while the effect of nonextensivity appears only through higher-order corrections. This observation is consistent with previous studies of nonextensive Maxwell–Boltzmann statistics, where the thermodynamic curvature was shown to remain identically zero over a range of $q$ values, indicating that nonextensivity by itself does not generate thermodynamic interactions \cite{adli2019condensation}.
\section{Applications for Pseudo-Exponential distribution}

To illustrate the applicability of the proposed pseudo-exponential distribution, two real data sets were analyzed. The first data set consists of the absolute difference between the WTI crude oil price and its 100-day moving average, using daily price data from 1986/01/02 to 2017/07/03. The second data set is based on the absolute difference between the Henry Hub Natural Gas (HHN) spot price and its 100-day moving average, using daily data from 1997/06/30 to 2017/06/30.

The model parameters were estimated by the maximum likelihood method using the statistical software R (version 3.4.1) with the nlminb package. According to the Akaike Information Criterion (AIC) and the Bayesian Information Criterion (BIC), the proposed pseudo-exponential distribution provided a better fit to both data sets than the classical exponential distribution and the other competing distributions considered\cite{mehri2019pseudo}.

\section{Applications for Pseudo-Tsallis distribution}

\begin{figure}[htbp]
	\centering
	\includegraphics[width=7.5 cm]{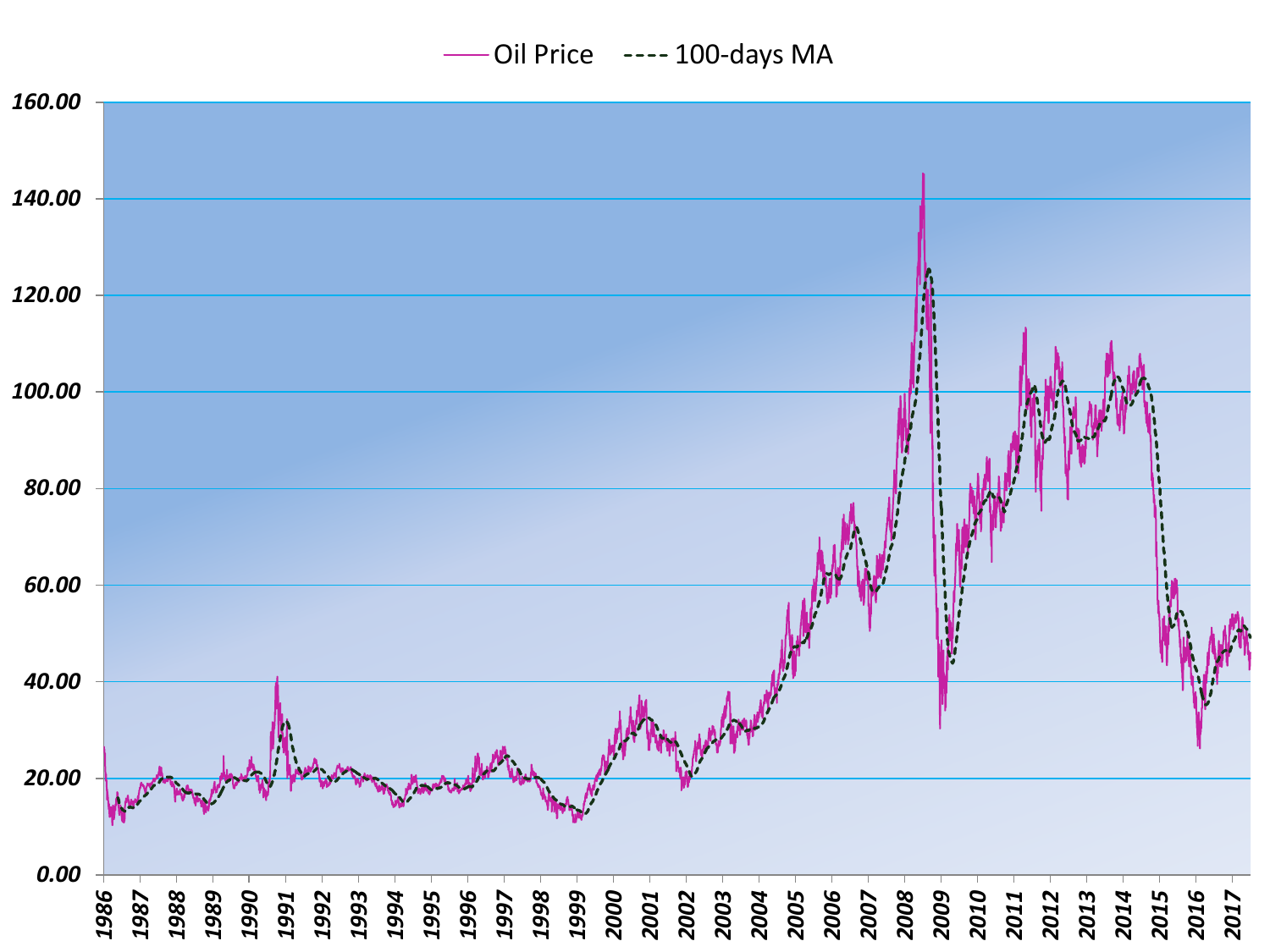}
	\caption{ The WTI crude oil, Oklahoma, dollars per barrel, daily price (full
		curve), and its 100 days moving average (dashed curve), in period 1986/01/02
		- 2017/07/03. }
	\label{Dec1}
\end{figure}

\begin{figure}[htbp]
\centering
\includegraphics[width=7.5 cm]{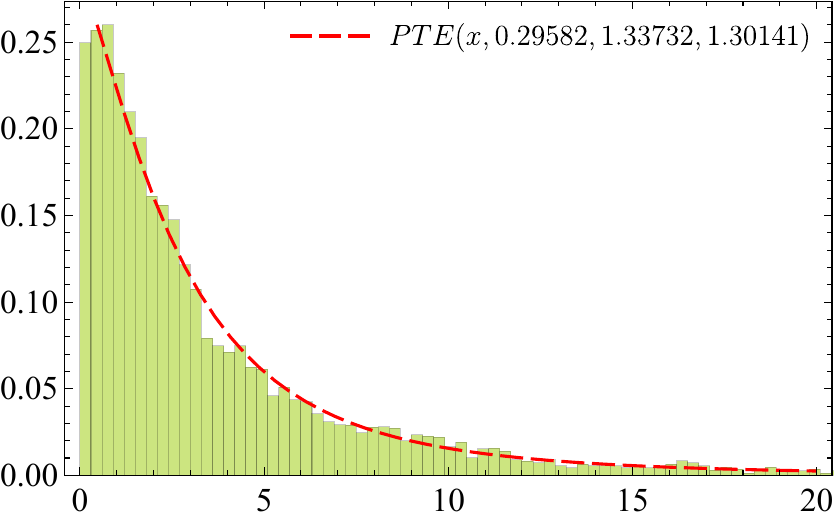}
\caption{ {The distribution of absolute difference of WTI crude oil, from its last 100 days moving average. The histogram obtained from WTI crude oil daily price in period of 1986/01/02 to 2017/07/03. } }
\label{Dec2}
\end{figure}

To describe the application of the proposed model, a real
dataset obtained from WTI crude oil, Oklahoma, dollars per barrel, daily
price, and its 100 days moving average, in period from 1986/01/02 to
2017/07/03 is examined (see Figure \ref{Dec1}). We work on the data of the
absolute difference of WTI crude oil price from its last 100 days moving
average in period of 1986 to 2017, described in Figure \ref{Dec2}.

Figure \ref{Dec2} shows the fitted curve of the pseudo-nonextensive Tsallis
distribution on the data. Throughout this section, we also used the
statistical software R version 3.4.1 with the package nlminb for estimating
the parameters. According to AIC and BIC, Table \ref{tab:FitExp} shows that the $%
PTE\left( \lambda ,q,\alpha \right) $ (pseudo-nonextensive Tsallis
distribution) better than other distributions.

\begin{table*}[t]
	\caption{\label{tab:FitExp}   
		Results of fitting different statistical models to WTI crude oil data.}
	\begin{ruledtabular}
		\begin{tabular}{lcccccc}
			Distribution & $\alpha$ & $\lambda$ & $q$ &
			$-\log \mathcal{L}$ & AIC & BIC \\
			\hline
			
			$\begin{array}{@{}l@{}}E(\lambda)\\
			\text{q-Tsallis exponential}
			\end{array}$
			&
			--
			&
			0.4091779
			&
			1.2056977
			&
			18704.36
			&
			37412.7
			&
			37426.7
			\\[12pt]
			
			$\begin{array}{@{}l@{}}TQG(\mu,\beta,q)\\
			\text{Tsallis-$q$-Gaussian}
			\end{array}$
			&
			$\mu=4.1527$
			&
			$\beta=2.66001$
			&
			1.48254
			&
			22549.2
			&
			45104.5
			&
			45125.4
			\\[12pt]
			
			$\begin{array}{@{}l@{}}E(\lambda)\\
			\text{Exponential}
			\end{array}$
			&
			--
			&
			0.240807
			&
			--
			&
			19024.07
			&
			38050.15
			&
			38057.12
			\\[12pt]
			
			$\begin{array}{@{}l@{}}GE(\alpha,\lambda)\\
			\text{G-Exponential}
			\end{array}$
			&
			0.6845109
			&
			0.6308010
			&
			--
			&
			18771.6
			&
			37547.2
			&
			37561.14
			\\[12pt]
			
			$\begin{array}{@{}l@{}}WE(\alpha,\lambda)\\
			\text{Weibull}
			\end{array}$
			&
			0.863852
			&
			3.82183
			&
			--
			&
			18856.49
			&
			37716.97
			&
			37730.91
			\\[12pt]
			
			$\begin{array}{@{}l@{}}N(\alpha,\lambda^2)\\
			\text{Normal}
			\end{array}$
			&
			4.1527
			&
			5.7462
			&
			--
			&
			24861.53
			&
			49727.06
			&
			49740.99
			\\[12pt]
			
			$\begin{array}{@{}l@{}}L(\alpha,\lambda)\\
			\text{Laplace}
			\end{array}$
			&
			2.30213
			&
			3.52587
			&
			--
			&
			22256.2
			&
			44516.4
			&
			44530.34
			\\[12pt]
			
			$\begin{array}{@{}l@{}}PTE(\lambda,q,\alpha)\\
			\text{pseudo-}q\text{-Tsallis exponential}
			\end{array}$
			&
			1.3014090
			&
			0.2958161
			&
			1.3373196
			&
			18692.02
			&
			\textbf{37390.0}
			&
			\textbf{37410.9}
			\\
			
		\end{tabular}
	\end{ruledtabular}
\end{table*}

\section{Conclusion}
In this paper, we have presented a systematic framework for constructing pseudo-generalized distributions using the formalism of $g$-calculus. By employing a monotone generator function to deform standard algebraic and exponential structures, we have shown that this approach provides a robust mechanism to generate a family of distributions that extend both classical and nonextensive statistics. The strength of this construction lies in its consistency across different levels of statistical description, including both probability density and cumulative distribution functions.

Our investigation into the thermodynamic geometry of these pseudo-distributions yielded critical insights into their structural properties. Through a perturbative analysis in the neighborhood of the classical limit, we demonstrated that the thermodynamic scalar curvature is primarily dictated by the deformation parameter of the $g$-calculus generator. Notably, the nonextensivity parameter remains decoupled from the scalar curvature at the leading order, revealing a clear hierarchy in how different deformation mechanisms influence the geometry of the equilibrium manifold.

The practical utility of the proposed model was further validated through an empirical application to high-frequency fluctuations in West Texas Intermediate crude oil prices. By analyzing the absolute deviations of these prices from their moving averages, we demonstrated that our $g$-calculus-based model provides a superior fit to the observed data compared to standard benchmarks. The ability of the model to effectively capture the heavy-tailed behavior inherent in these financial fluctuations underscores its potential for physical and socio-economic modeling.

In summary, the $g$-calculus approach offers a unified and flexible mathematical route to explore deformed statistical laws. Future research could extend this geometric analysis to higher-order terms in the expansion, examine alternative classes of generator functions, and apply this formalism to multi-variable systems, such as those found in quantum gravity or complex networks, where scale-modified mechanics play a central role.

\appendix
\begin{widetext}
\section{Relations for the pseudo-Exponential distribution}

In this appendix, we collect the analytical expressions obtained for the particle number, the internal energy, and the thermodynamic curvature associated with the pseudo-exponential distribution.
For the pseudo-generalized distribution constructed from the probability density function (PDF), the first–order corrected quantities $N_1$, $U_1$, and $R_1$ are given by

\begin{equation}
\begin{aligned}
N_1&= \frac{\sqrt{\pi} z }{4 \beta^{3/2}}
\Bigg[
-6 + 3\gamma -2\gamma_E + 3\ln 4
+ \frac{6\sqrt{\pi}\sqrt{-\ln z}}{z}
\\[2mm]
&\quad
+ \frac{3\pi \operatorname{erf}(\sqrt{\ln z})\sqrt{\ln z}}
{\sqrt{-\ln z}}
+ (4 - 2\gamma - 2\ln4)\ln z
\\[2mm]
&\quad
+ 2\,{}_2F_2\!\left(1,1;-\tfrac12,2;-\ln z\right)\ln z
- 4\,{}_2F_2\!\left(1,1;\tfrac12,2;-\ln z\right)(\ln z)^2
\\[2mm]
&\quad
+ 2\pi\,\operatorname{erf}(\sqrt{\ln z})\sqrt{(-\ln z)^2}
\Bigg],
\end{aligned}
\end{equation}

\begin{equation}
\begin{aligned}
U_1 &= \frac{\sqrt{\pi} z}{8 \beta^{5/2}} 
\Bigg[
-40 + 15\gamma -6\gamma_E + 15\ln 4 + \frac{30\sqrt{\pi}\sqrt{-\ln z}}{z} - \frac{8\sqrt{\pi}(-\ln z)^{3/2}}{z}
\\[2mm]
&\quad
+ \frac{15\pi\operatorname{erf}(\sqrt{\ln z})\sqrt{\ln z}}{\sqrt{-\ln z}} + (16 - 6\gamma- 6\ln 4) \ln z 
\\[2mm]
&\quad
+ 6 \, {}_2F_2(1,1; -\tfrac{3}{2}, 2; -\ln z) \ln z - 4 \, {}_2F_2(1,1; -\tfrac{1}{2}, 2; -\ln z) (\ln z)^2 
\\[2mm]
&\quad
+ 6\pi\operatorname{erf}(\sqrt{\ln z})\sqrt{(-\ln z)^2} 
\Bigg],
\end{aligned}
\end{equation}

\begin{equation}
\begin{split}
R_{1} =& -\frac{5 \beta^{3/2}}{12 \sqrt{\pi} z^2 \sqrt{-\ln z} \sqrt{-(\ln z)^2}} 
\bigg[ 9\sqrt{\pi}\sqrt{-\ln z}\sqrt{\ln z} + 12\pi z \operatorname{erf}(\sqrt{\ln z})\sqrt{-\ln z} \ln z \\
&+ 6\sqrt{\pi}\sqrt{-\ln z} (\ln z)^{3/2} - 9\sqrt{\pi}\sqrt{-(\ln z)^2} + 8z \sqrt{-\ln z}\sqrt{-(\ln z)^2} \\
&- 12\pi z \operatorname{erf}(\sqrt{\ln z})\sqrt{\ln z}\sqrt{-(\ln z)^2} - 12\sqrt{\pi} \operatorname{erfi}(\sqrt{\ln z})\sqrt{-\ln z}\sqrt{\ln z}\sqrt{-(\ln z)^2} \\
&+ 6\sqrt{\pi} \ln z \sqrt{-(\ln z)^2} - 8z \sqrt{-\ln z} \ln z \sqrt{-(\ln z)^2} \\
&- 12z {}_2F_2(1,1;-\tfrac{3}{2},2;-\ln z) \sqrt{-\ln z} \ln z \sqrt{-(\ln z)^2} \\
&+ 24z {}_2F_2(1,1;-\tfrac{1}{2},2;-\ln z) \sqrt{-\ln z} \ln z \sqrt{-(\ln z)^2} \\
&- 12z {}_2F_2(1,1;\tfrac{1}{2},2;-\ln z) \sqrt{-\ln z} \ln z \sqrt{-(\ln z)^2} \\
&- 8\sqrt{\pi} (\ln z)^2 \sqrt{-(\ln z)^2} + 8z {}_2F_2(1,1;-\tfrac{1}{2},2;-\ln z) \sqrt{-\ln z} (\ln z)^2 \sqrt{-(\ln z)^2} \\
&- 24z {}_2F_2(1,1;\tfrac{1}{2},2;-\ln z) \sqrt{-\ln z} (\ln z)^2 \sqrt{-(\ln z)^2} \bigg].
\end{split}
\end{equation}

Similarly, for the pseudo-exponential distribution derived from the pseudo cumulative distribution function (CDF), the corresponding first–order corrections take the forms
\begin{equation}
\begin{aligned}
N_{1} &=\frac{z}{\beta^{3/2}}
\Bigg[
\left( -\frac{\pi \sqrt{-\ln z}}{z} + \frac{\pi^{3/2} \operatorname{erf}(\sqrt{\ln z}) \sqrt{-\ln z}}{2 \sqrt{\ln z}} \right)
\\[2mm]
&\quad - \frac{\sqrt{\pi}}{2} \left( -2 + \gamma + \ln 4 + 2 \, {}_2F_2(1,1; \tfrac{1}{2}, 2; -\ln z) \ln z \right)
\\[2mm]
&\quad - \frac{\sqrt{\pi}}{4\sqrt{\ln z}} \Bigg( 
3 \pi z \operatorname{erf}(\sqrt{\ln z}) \sqrt{-\ln z} + 2 \pi z \operatorname{erf}(\sqrt{\ln z}) (-\ln z)^{3/2} + 8 z \sqrt{\ln z}
\\[1mm]
&\quad - 3 \gamma z \sqrt{\ln z} - 3 z \ln 4 \sqrt{\ln z} - 4 z (\ln z)^{3/2} + 2 \gamma z (\ln z)^{3/2}
\\[1mm]
&\quad - 2 z \, {}_2F_2(1,1; -\tfrac{1}{2}, 2; -\ln z) (\ln z)^{3/2} + z \ln 16 (\ln z)^{3/2} 
\\[1mm]
&\quad + 4 z \, {}_2F_2(1,1; \tfrac{1}{2}, 2; -\ln z) (\ln z)^{5/2}  - 6 \sqrt{\pi} \sqrt{(-\ln z)^2} \Bigg)
\\[2mm]
&\quad + z \Gamma(3/2) \sum_{n=1}^{\infty} \frac{z^n}{n (n+1)^{3/2}}\Bigg],
\end{aligned}
\end{equation}

\begin{equation}
\begin{aligned}
U_{1} &=\frac{\sqrt{\pi} z}{4 \beta^{5/2}} 
\Bigg[
\left( 8 - 3 \gamma - 3 \ln 4 - \frac{6 \sqrt{\pi} \sqrt{-\ln z}}{z} + \frac{4 \sqrt{\pi} (-\ln z)^{3/2}}{z} \right)
\\[2mm]
&\quad + \left( \frac{3 \pi \operatorname{erf}(\sqrt{\ln z}) \sqrt{-\ln z}}{\sqrt{\ln z}} - 2 \, {}_2F_2(1,1; -\tfrac{1}{2}, 2; -\ln z) \ln z \right)
\\[2mm]
&\quad - \frac{1}{8 \sqrt{\ln z}} \Bigg( 
15 \pi z \operatorname{erf}(\sqrt{\ln z}) \sqrt{-\ln z} + 6 \pi z \operatorname{erf}(\sqrt{\ln z}) (-\ln z)^{3/2} 
\\[1mm]
&\quad + 46 z \sqrt{\ln z} - 15 \gamma z \sqrt{\ln z} - 15 z \ln 4 \sqrt{\ln z} + 8 \sqrt{\pi} (-\ln z)^{3/2} \sqrt{\ln z} 
\\[1mm]
&\quad - 16 z (\ln z)^{3/2} + 6 \gamma z (\ln z)^{3/2} - 6 z \, {}_2F_2(1,1; -\tfrac{3}{2}, 2; -\ln z) (\ln z)^{3/2}
\\[1mm]
&\quad  + 6 z \ln 4 (\ln z)^{3/2} + 4 z \, {}_2F_2(1,1; -\tfrac{1}{2}, 2; -\ln z) (\ln z)^{5/2} - 30 \sqrt{\pi} \sqrt{(-\ln z)^2} \Bigg)
\\[2mm]
&\quad + z \Gamma(5/2) \sum_{n=1}^{\infty} \frac{z^n}{n (n+1)^{5/2}}\Bigg],
\end{aligned}
\end{equation}

\begin{equation}
\begin{aligned}
R_{1} &= -\frac{5\beta^{3/2}}{24\sqrt{\pi}z^{2}\ln^{2}z\sqrt{-\ln^{2}z}}
\Bigg[
-9\sqrt{\pi}\ln^{3/2}z
+36\sqrt{\pi}\ln^{5/2}z
-60\sqrt{\pi}\ln^{7/2}z
\\
&\quad
+\left(
-9\sqrt{\pi}
+24\sqrt{\pi}\ln z
-36\sqrt{\pi}\ln^{2}z
\right)
\sqrt{-\ln z}\sqrt{-\ln^{2}z}
\\
&\quad
+\operatorname{erfi}(\sqrt{\ln z})\sqrt{-\ln^{2}z}
\left(
-12\sqrt{\pi}\ln^{3/2}z
+24\sqrt{\pi}\ln^{5/2}z
-48\sqrt{\pi}\ln^{7/2}z
\right)
\\
&\quad
+z^{2}\ln^{2}z\sqrt{-\ln^{2}z}
\left[
-12\,\Phi\!\left(z,\tfrac{1}{2},2\right)
+24\,\Phi\!\left(z,\tfrac{3}{2},2\right)
-36\,\Phi\!\left(z,\tfrac{5}{2},2\right)
\right]
\\
&\quad
+z\ln^{2}z\sqrt{-\ln^{2}z}
\left(
-24
+12\ln4
-6\ln16
\right)
\\
&\quad
+z\ln^{3}z\sqrt{-\ln^{2}z}
\left(
32
+24\ln4
-12\ln16
\right)
\\
&\quad
+z\ln^{3}z\sqrt{-\ln^{2}z}
\Big[
-24\,{}_2F_2\!\left(1,1;-\tfrac{3}{2},2;-\ln z\right)
\\
&\qquad\qquad\qquad\quad
+64\,{}_2F_2\!\left(1,1;-\tfrac{1}{2},2;-\ln z\right)
\\
&\qquad\qquad\qquad\quad
-72\,{}_2F_2\!\left(1,1;\tfrac12,2;-\ln z\right)
\Big]
\\
&\quad
+z\ln^{4}z\sqrt{-\ln^{2}z}
\Big[
16\,{}_2F_2\!\left(1,1;-\tfrac12,2;-\ln z\right)
\\
&\qquad\qquad\qquad\quad
-48\,{}_2F_2\!\left(1,1;\tfrac12,2;-\ln z\right)
\Big]
\\
&\quad
-24z\ln^{2}z\sqrt{-\ln^{2}z}
\sum_{n=1}^{\infty}\frac{z^{n}}{n(1+n)^{5/2}}
\\
&\quad
+24z\ln^{2}z\sqrt{-\ln^{2}z}
\sum_{n=1}^{\infty}\frac{z^{n}}{n(1+n)^{3/2}}
\Bigg].
\end{aligned}
\end{equation}

These expressions provide the complete first–order contributions to the particle number, internal energy, and curvature for both constructions and are used throughout the main text.

\bigskip

Here $\gamma_E$ is the Euler--Mascheroni constant, 
$\operatorname{erf}(x)$ denotes the error function, 
$\operatorname{erfi}(x)$ denotes the imaginary error function, 
${}_pF_q$ represents the generalized hypergeometric function, 
and $\Phi(z,s,a)$ stands for the Lerch transcendent.\\

\section{Relations for the pseudo-Tsallis distribution}
We list the analytical expressions for the particle number, the internal energy, and the thermodynamic curvature associated with the pseudo-Tsallis distribution.
For the pseudo-Tsallis distribution constructed from the probability density function (PDF), the first-order corrected quantities 
$N_{\alpha}(z,\beta)$, $N_{q}(z,\beta)$, $U_{\alpha}(z,\beta)$, $U_{q}(z,\beta)$, and $R_{\alpha}(z,\beta)$ are written as
\begin{equation}
\begin{aligned}
N_{\alpha}&=
\frac{\sqrt{\pi}z}{16\beta^{3/2}}
\Bigg[
-20+12\gamma_E+12\ln4
+\frac{24\sqrt{\pi}\sqrt{-\ln z}}{z}
\\
&\qquad
+\frac{12\pi\operatorname{erf}(\sqrt{\ln z})\sqrt{\ln z}}
{\sqrt{-\ln z}}
+\left(8-8\gamma_E-8\ln4\right)\ln z
\\
&\qquad
+8\,{}_2F_2\!\left(1,1;-\frac12,2;-\ln z\right)\ln z
-16\,{}_2F_2\!\left(1,1;\frac12,2;-\ln z\right)(\ln z)^2
\\
&\qquad
+8\pi\operatorname{erf}(\sqrt{\ln z})
\sqrt{(-\ln z)^2}
\Bigg],
\\[2mm]
N_q&=
\frac{\sqrt{\pi}z}{16\beta^{3/2}}
\left[
7-12\ln z+4(\ln z)^2
\right].
\end{aligned}
\end{equation}

\begin{equation}
\begin{aligned}
U_{\alpha}&=
\frac{\sqrt{\pi}z}{32\beta^{5/2}}
\Bigg[
-124+60\gamma_E+60\ln4
+\frac{120\sqrt{\pi}\sqrt{-\ln z}}{z}
-\frac{32\sqrt{\pi}(-\ln z)^{3/2}}{z}
\\
&\qquad
+\frac{60\pi\operatorname{erf}(\sqrt{\ln z})\sqrt{\ln z}}
{\sqrt{-\ln z}}
+\left(40-24\gamma_E-24\ln4\right)\ln z
\\
&\qquad
+24\,{}_2F_2\!\left(1,1;-\frac32,2;-\ln z\right)\ln z
-16\,{}_2F_2\!\left(1,1;-\frac12,2;-\ln z\right)(\ln z)^2
\\
&\qquad
+24\pi\operatorname{erf}(\sqrt{\ln z})
\sqrt{(-\ln z)^2}
\Bigg],
\\[2mm]
U_q&=
\frac{\sqrt{\pi}z}{32\beta^{5/2}}
\left[
81-60\ln z+12(\ln z)^2
\right].
\end{aligned}
\end{equation}

\begin{equation}
\begin{aligned}
R_{\alpha}&=
-\frac{10\beta^{3/2}}{3\sqrt{\pi}\,z}
+\frac{5\beta^{3/2}\sqrt{-\ln z}}{2z^2}
-\frac{15\beta^{3/2}\sqrt{-\ln z}}{4z^2\ln z}
\\[2mm]
&
-\frac{5\sqrt{\pi}\beta^{3/2}
	\operatorname{erf}(\sqrt{\ln z})\sqrt{-\ln z}}
{z\sqrt{\ln z}}
+\frac{5\beta^{3/2}
	\operatorname{erfi}(\sqrt{\ln z})\sqrt{\ln z}}
{z^2}
\\[2mm]
&
+\frac{10\beta^{3/2}\ln z}{3\sqrt{\pi}\,z}
+\frac{5\beta^{3/2}}{\sqrt{\pi}\,z}
\,{}_2F_2\!\left(1,1;-\frac32,2;-\ln z\right)\ln z
\\[2mm]
&
-\frac{10\beta^{3/2}}{\sqrt{\pi}\,z}
\,{}_2F_2\!\left(1,1;-\frac12,2;-\ln z\right)\ln z
+\frac{5\beta^{3/2}}{\sqrt{\pi}\,z}
\,{}_2F_2\!\left(1,1;\frac12,2;-\ln z\right)\ln z
\\[2mm]
&
-\frac{10\beta^{3/2}\sqrt{-\ln z}\,\ln z}{3z^2}
-\frac{10\beta^{3/2}}{3\sqrt{\pi}\,z}
\,{}_2F_2\!\left(1,1;-\frac12,2;-\ln z\right)(\ln z)^2
\\[2mm]
&
+\frac{10\beta^{3/2}}{\sqrt{\pi}\,z}
\,{}_2F_2\!\left(1,1;\frac12,2;-\ln z\right)(\ln z)^2
\\[2mm]
&
+\frac{15\beta^{3/2}\sqrt{(-\ln z)^2}}
{4z^2(\ln z)^{3/2}}
+\frac{5\sqrt{\pi}\beta^{3/2}
	\operatorname{erf}(\sqrt{\ln z})
	\sqrt{(-\ln z)^2}}
{z\ln z}
+\frac{5\beta^{3/2}\sqrt{(-\ln z)^2}}
{2z^2\sqrt{\ln z}}.
\end{aligned}
\end{equation}

Likewise, for the pseudo-Tsallis distribution defined through the pseudo cumulative distribution function (CDF), the corresponding first-order corrections are obtained as

\begin{equation}
\begin{aligned}
N_{\alpha}&=
\frac{\sqrt{\pi}}{4\beta^{3/2}\sqrt{\ln z}}
\Bigg[
-\pi z\,\operatorname{erf}(\sqrt{\ln z})\sqrt{-\ln z}
\\[2mm]
&\qquad
-2\pi z\,\operatorname{erf}(\sqrt{\ln z})(-\ln z)^{3/2}
-4z\sqrt{\ln z}
+\gamma_E z\sqrt{\ln z}
+z\ln4\,\sqrt{\ln z}
\\[2mm]
&\qquad
+4z(\ln z)^{3/2}
-2\gamma_E z(\ln z)^{3/2}
+2z\,{}_2F_2\!\left(1,1;-\frac12,2;-\ln z\right)(\ln z)^{3/2}
\\[2mm]
&\qquad
-2z\ln4\,(\ln z)^{3/2}
+2\sqrt{\pi}\sqrt{(-\ln z)^2}
\\[2mm]
&\qquad
-4z\,{}_2F_2\!\left(1,1;\frac12,2;-\ln z\right)
(\ln z)^{3/2}(1+\ln z)
\Bigg]
\\[2mm]
&\qquad
+\frac{\sqrt{\pi}\,z}{2\beta^{3/2}}
\sum_{n=1}^{\infty}
\frac{z^n}{n(n+1)^{3/2}},
\\[2mm]
N_q
&=
\frac{\sqrt{\pi}\,z}{16\beta^{3/2}}
\left[
7-12\ln z+4(\ln z)^2
\right].
\end{aligned}
\end{equation}

\begin{equation}
\begin{aligned}
U_{\alpha}&=
\frac{\sqrt{\pi}}{8\beta^{5/2}\sqrt{\ln z}}
\Bigg[
-9\pi z\,\operatorname{erf}(\sqrt{\ln z})\sqrt{-\ln z}
\\[2mm]
&\qquad
-6\pi z\,\operatorname{erf}(\sqrt{\ln z})(-\ln z)^{3/2}
-30z\sqrt{\ln z}
+9\gamma_E z\sqrt{\ln z}
+9z\ln4\,\sqrt{\ln z}
\\[2mm]
&\qquad
+16z(\ln z)^{3/2}
-6\gamma_E z(\ln z)^{3/2}
+6z\,{}_2F_2\!\left(1,1;-\frac32,2;-\ln z\right)(\ln z)^{3/2}
\\[2mm]
&\qquad
-6z\ln4\,(\ln z)^{3/2}
+18\sqrt{\pi}\sqrt{(-\ln z)^2}
\\
&\qquad
-4z\,{}_2F_2\!\left(1,1;-\frac12,2;-\ln z\right)
(\ln z)^{3/2}(1+\ln z)
\Bigg]
\\[2mm]
&\qquad
+\frac{3\sqrt{\pi}\,z}{4\beta^{5/2}}
\sum_{n=1}^{\infty}
\frac{z^n}{n(n+1)^{5/2}},
\\[2mm]
U_q
&=
\frac{3\sqrt{\pi}\,z}{32\beta^{5/2}}
\left[
27-20\ln z+4(\ln z)^2
\right].
\end{aligned}
\end{equation}

\begin{equation}
\begin{aligned}
R_{\alpha}&=
\frac{5\beta^{3/2}}{\sqrt{\pi}z}
+\frac{5\beta^{3/2}}{2\sqrt{\pi}}
\Phi\!\left(z,\frac12,2\right)
-\frac{5\beta^{3/2}}{\sqrt{\pi}}
\Phi\!\left(z,\frac32,2\right)
+\frac{15\beta^{3/2}}{2\sqrt{\pi}}
\Phi\!\left(z,\frac52,2\right)
\\[2mm]
&\qquad
+\frac{5\beta^{3/2}\sqrt{-\ln z}}{2z^2}
+\frac{5\beta^{3/2}\sqrt{-\ln z}}
{8z^2(\ln z)^2}
-\frac{5\beta^{3/2}\sqrt{-\ln z}}
{z^2\ln z}
\\[2mm]
&\qquad
+\frac{5\beta^{3/2}\operatorname{erfi}(\sqrt{\ln z})}
{2z^2\sqrt{\ln z}}
-\frac{5\beta^{3/2}\operatorname{erfi}(\sqrt{\ln z})
	\sqrt{\ln z}}
{z^2}
\\[2mm]
&\qquad
-\frac{20\beta^{3/2}\ln z}
{3\sqrt{\pi}z}
+\frac{5\beta^{3/2}}
{\sqrt{\pi}z}
{}_2F_2\!\left(1,1;-\frac32,2;-\ln z\right)\ln z
\\[2mm]
&\qquad
-\frac{40\beta^{3/2}}
{3\sqrt{\pi}z}
{}_2F_2\!\left(1,1;-\frac12,2;-\ln z\right)\ln z
+\frac{15\beta^{3/2}}
{\sqrt{\pi}z}
{}_2F_2\!\left(1,1;\frac12,2;-\ln z\right)\ln z
\\[2mm]
&\qquad
+\frac{10\beta^{3/2}\operatorname{erfi}(\sqrt{\ln z})
	(\ln z)^{3/2}}
{z^2}
\\[2mm]
&\qquad
-\frac{10\beta^{3/2}}
{3\sqrt{\pi}z}
{}_2F_2\!\left(1,1;-\frac12,2;-\ln z\right)
(\ln z)^2
\\[2mm]
&\qquad
+\frac{10\beta^{3/2}}
{\sqrt{\pi}z}
{}_2F_2\!\left(1,1;\frac12,2;-\ln z\right)
(\ln z)^2
\\[2mm]
&\qquad
-\frac{5\beta^{3/2}\sqrt{(-\ln z)^2}}
{8z^2(\ln z)^{5/2}}
+\frac{15\beta^{3/2}\sqrt{(-\ln z)^2}}
{2z^2(\ln z)^{3/2}}
-\frac{15\beta^{3/2}\sqrt{(-\ln z)^2}}
{2z^2\sqrt{\ln z}}
\\[2mm]
&\qquad
+\frac{5\beta^{3/2}}
{\sqrt{\pi}z}
\sum_{n=1}^{\infty}
\frac{z^n}{n(n+1)^{5/2}}
-\frac{5\beta^{3/2}}
{\sqrt{\pi}z}
\sum_{n=1}^{\infty}
\frac{z^n}{n(n+1)^{3/2}}.
\end{aligned}
\end{equation}

\end{widetext}







\bibliographystyle{apsrev4-2}
\bibliography{References}

\end{document}